# Reactive infiltration instability amplifies the difference between geometric and reactive surface areas in natural porous materials

*Y. Yang\*, S. Bruns, S. L. S. Stipp* and *H. O. Sørensen*

*Nano-Science Center, Department of Chemistry, University of Copenhagen, Universitetsparken 5, DK-2100 Copenhagen, Denmark*

**ABSTRACT** Reactive infiltration instability (RII) drives the development of many natural and engineered flow systems. These are encountered e.g. in hydraulic fracturing, geologic carbon storage and well stimulation in enhanced oil recovery. The surface area of the rocks changes as the pore structure evolves. We combined a reactor network model with grey scale tomography to seek the morphological interpretation for differences among geometric, reactive and apparent surface areas of dissolving natural porous materials. The approach allowed us to delineate the experimentally convoluted variables and study independently the effects of initial geometry and macroscopic flowrate. Simulations based on North Sea chalk microstructure showed that geometric surface not only serves as the interface for water-rock interactions but also represents the regional transport heterogeneities that can be amplified indefinitely by dissolutive percolation. Hence, RII leads to channelization of the solid matrix, which results in fluid focusing and an increase in geometric surface area. Fluid focusing reduces the reactive surface area and the residence time of reactants, both of which amplify the differences in question, i.e. they are self-supporting. Our results also suggested that the growing and merging of microchannels near the fluid entrance leads to the macroscopic "fast initial dissolution" of chemically homogeneous materials.

**INTRODUCTION**

Reactive infiltration instability (RII) is the geometric instability of a propagating reaction front induced by a positive coupling between the regional permeability of a porous medium and the chemical affinity of reactions that modify the solid volume.[1] This instability amplifies regional heterogeneities in the transport and reactive properties of porous media and drives the morphological evolution of various flow systems (e.g., Figure S1).[2,3] An improved understanding of how these pressure driven geochemical systems evolve is particularly important for several research scenarios of energy and environmental concern. For example, in geologic carbon storage (GCS), reservoir structures change near the injection wells in response to elevated regional pressure, acidified brine and as a consequence mineral dissolution and reprecipitation occurs.[4-10] Small defects in caprocks can also be amplified by RII, leading to the failure of sealing structures.[11,12] In hydraulic fracturing[13] and well stimulation in enhanced oil recovery (EOR),[14] infiltration instability is introduced deliberately to induce morphological changes in the reservoir.

Quantitative analysis of RII-dominated geochemical systems is still in its infancy. The challenges are threefold. First, the evolution of such systems can be triggered by very small regional features,[15,16] and thus high precision characterization of the initial geometry is needed.[17,18] Second, the many inherently coupled processes in a natural RII system are virtually impossible to delineate experimentally.[19] For example, solution chemistry and operational conditions presumably affect the structural evolution of natural porous materials during water-rock interactions. However, each percolation experiment would destroy a sample, and no two natural samples have the identical internal structure and chemical heterogeneity. One thus cannot start with the same initial condition and vary the solution chemistry alone to see its effects. Such investigation can be done through numerical simulations,[20-23] but the latter itself embodies the third difficulty. Quantitative study of RII requires a mathematical scheme that predicts the morphological development unambiguously, so the simulation can be compared directly with experimental observations. However, tackling free boundary problems of partial differential equations – a conventional way of handling fluid flow in a morphing confinement – usually requires formidable computational power.[22, 24] Moreover, much regional heterogeneity recorded by structural characterisation techniques is removed by image segmentation before topological data are imported into a numerical simulation.[25-27] For example, lumping voxels of similar grey levels into one phase removes immediately the spatial variations in material density that is reflected by intensity differences among voxels. This reduction of information has demonstrated differing effects depending on the rock textures.[28] Such over-simplification of initial geometry may lead to significant uncertainties if RII is to govern the morphology development.

Among the many geochemistry problems that could benefit from a dynamic insight into the RII-induced microstructure development, most fascinating is the discrepancy between geometric and reactive surface areas of geological materials.[21, 29] The geometric surface reflects the amplitude and frequency of the spatial variations of material density [30] whereas the reactive surface is the geometric surface on which heterogeneous chemical reactions are taking place at a given instant.[31] The poor match of the two has been considered a main cause of the long lasting "enigma" of geochemistry: the laboratory-field rate discrepancy.[32] This apparent discrepancy in rate measurement undermines the scientific justifications behind numerous quantitative models of earth sciences.[21] Plausible explanations fall into three categories. First, phase contacting: Discrepancies arise when parts of geometric surface are not in contact with fluid. This may occur when the pressure field does not allow the complete rinse of solid surface,[33] or when some internal pore structures are physically isolated from interactions with the environment.[34] Second, uneven distribution of fluid reactivity: The rates of geochemical reactions depend on the chemical affinity and thus the reactant concentration.[21, 24] Momentum and mass transfer in complex microstructures usually lead to dramatic spatial variations of fluid composition.[35, 36] Third, uneven distribution of solid reactivity: This may stem from mineralogical heterogeneity, surface modification by secondary mineral reprecipitation,[37, 38] or more generally the diversity of the atomic structures of rock surfaces as assemblages of chemical bonds, including the effects from crystalline orientation and bond truncation.[39, 40] None of these are mutually exclusive yet most interpretations remain qualitative because of the difficulty in delineation. Thus questions remain: How much can each of these mechanisms contribute to the overall discrepancy? Will the pores initially isolated from the flow field remain isolated? Will the discrepancy remain if the solid surface is chemically homogeneous and no precipitation occurs? An unequivocal answer to these questions calls for a systematic approach to put results of experimentation, characterisation, modelling and numerical simulation into a coherent picture.

Here we present a reactor network model built on grey scale tomographic characterisations of porous media. The microstructure of natural chalk was used as the model geometry because of chalk reservoir's importance in drinking water safety, EOR and GCS in Scandinavia.[41-43] The model was designed to quantify emergent phenomena in porous media and allowed the delineation of influences from experimentally convoluted factors. We studied the effects of microstructural evolution on surface areas defined in three representative ways: geometric, reactive and apparent. The simulations revealed morphological changes as the pore-scale origins of various macroscopic behaviours of porous media, including surface area discrepancies, macroscopic flowrate dependences and the fast initial dissolution of minerals.[44]

**MATERIALS AND METHODS**

Drill cuttings of Hod chalk (North Sea Basin), identified as sample HC #16, provided air dried chalk particles of ~500 μm diameter (B.E.T. surface area: ~7.0 m$^2$/g)[42] that were used for tomographic imaging to generate a model environment close to realistic reservoir conditions. Imaging was performed with the X-ray holotomography setup at the former ID22 beamline (29.49 keV) at the European Synchrotron Research Facility in Grenoble, France.[45] The data were reconstructed from 1999 radiographs (360° rotation, 0.5 s exposure) at 100 nm voxel resolution using the holotomography reconstruction method and processed as described in Bruns et al.[46, 47] The goal of the processing routine was to generate a greyscale volume image where variations in voxel intensity could be related to local material density, i.e. greyscale variations result from partial volume effects and not from signal blur, noise or artefacts. The salient points of the image processing routine were (i) background compensation by Fourier highpass filtering, (ii) ring artefacts were removed by polar transformation and median filtering[48], (iii) noise was reduced by iterative nonlocal means denoising [Bruns2015b], (iv) deconvolution under the assumption of a Gaussian point spread function and (v) transformation to voxel level porosity by linear interpolation between the average greyscale intensity of chalk and void phase. Since the average intensities of these phases are initially unknown, a seven phase Gaussian mixture model was used to identify the most likely intensities for chalk and void phase resulting in a macroscopic porosity value of 0.22 for the central volume of 1350x1350x1514 voxels, the largest volume fully inside the sample. From this reconstruction samples of 1.08 million voxel cubic regions were chosen randomly.

In the numerical simulations we modelled voxels as Continuously Stirred Tank Reactors (CSTRs). Each CSTR was connected to 6 neighbors by Plug Flow Reactors (PFRs). The volume of a voxel was thus divided into the 6 half PFRs and 1 CSTR based on the grey level of the voxel (Figures S2 to S4). CSTR and PFR represent two extremes of mass transfer: in CSTR reactants and products are completely mixed while in PFR no mixing occurs.[49] A mixing factor $\eta$, varying from 0 to 1, indicates the fraction of a voxel occupied by the CSTR. The advantage in using a reactor network to compute chemical conversions was that it only requires the mixing state and a reaction rate law as input, and thus circumvents the need for tracking specific momentum and mass transfer mechanisms. This convenience trades in precision – given the kinetics knowledge, the network model produced the upper and the lower bounds of chemical conversion (corresponding to $\eta = 1$ and 0). Although it was necessary to introduce this artificial parameter (mixing factor $\eta$) to specify sub-voxel mixing states, the value of $\eta$ did not affect the qualitative nature of the microstructural evolution. A detailed discussion of sub-voxel mixing effects is beyond the scope of this manuscript and will be presented separately. The $\eta$ value was set to 0.5 in all simulations in this study to ensure the presence of both mixing mechanisms. The porosity of each voxel was then updated according to the chemical conversions in its constituent reactors.

The pressure field was evaluated as the current distribution in the resistor network (Figure S3). We assumed that pressure only dropped in the PFRs and that it obeys Darcy's law

$$q = \left(\frac{P_{ref} L_{ref}}{\mu Q_{ref}}\right) \cdot \left(\frac{\partial \kappa}{\partial \varphi}\right) \cdot l_n \varphi^2 \cdot \Delta p, \qquad (1)$$

where $q$ represents the normalized volumetric flow rate, $\mu$ the viscosity of the reactive fluid [Pa·s], $\varphi^2$ the product of the porosities of the neighboring voxels connected by the PFR, $l_n$ the normalized voxel dimension, $p$ the normalized pressure, $\frac{\partial \kappa}{\partial \varphi}$ the first order Taylor coefficient of voxel level permeability [m$^2$] and $P_{ref}$, $L_{ref}$ and $Q_{ref}$ represent the reference pressure [Pa], length [m] and volumetric flowrate [m$^3$/s]. Further discussion regarding the dimensional analysis and the choice of the reference values can be found in the SI.

We employed a linear reversible reaction scheme $A_{solid} \leftrightarrow A_{solute}$ to describe the microstructure dissolution: $r_A = k_A \left( C_{A,eq} - C_A \right)$, where $r_A$ represents the mineral dissolution rate [mol·m$^{-2}$·s$^{-1}$], $k_A$ is the first order rate constant [m·s$^{-1}$], $C_{A,eq}$ is the equilibrium concentration of dissolved solid $A$ [mol·m$^{-3}$] and $C_A$ is the aqueous concentration of $A$ [mol·m$^{-3}$]. This scheme had the advantage of avoiding complications from "earliness of mixing"[50] (discussed in the SI) without losing the general chemical affinity dependence of a rate law.[51] The performance equations of PFR and CSTR were given by

$$-C_0 + e^{Da} \cdot C = 0 \quad (2)$$

and

$$-\sum_i q_i C_{0,i} + (1+Da) \cdot qC = 0, \quad (3)$$

where $C_0$ and $C$ represent normalized inlet and outlet reactant concentrations. The subscript $i$ indexes the number of CSTR inlets according to the pressure field. $Da$ is the first order Damköhler number[36] in which the regional geometric surface area is computed based on a summation of porosity differences over the 6 neighbor pairs of a voxel:

$$Da = \left( \frac{L_{ref}^2}{Q_{ref}} \right) \cdot k_A \cdot \eta \cdot l_n^2 \cdot \left( \frac{\sum_{i=1}^{6} |\nabla \varphi|_i}{q} \right). \quad (4)$$

A pseudo steady state of fluid distribution was assumed before the porosities of voxels were updated according to chemical conversions.[20] This assumption is justifiable when the chemical reactions in question are much slower than the establishment of the flow field.[20] The porosity change was then calculated as

$$\frac{d\varphi}{dt} = \left( \frac{Q_{ref}}{L_{ref}^3} \right) \cdot \left( \frac{M}{\rho} \right) \cdot \left( C_{A,eq} - C_{A,inj} \right) \cdot l_n^{-3} \cdot \sum_{i=1}^{7} q_i \left( C_{0,i} - C_i \right), \quad (5)$$

where $M$ and $\rho$ are the molar weight (g·mol$^{-1}$) and the density (g·L$^{-1}$) of the mineral and $C_{A,inj}$ is the injecting concentration of A. The geometric surface area ($SA_{geo}$, m$^2$) was computed based on the spatial variations of material density within the simulation domain

$$SA_{geo} = l^2 \cdot \iiint_V |\nabla \varphi| \cdot d\mathbf{r} \ . \qquad (6)$$

where $l$ is the voxel size [m]. Reactive surface area ($SA_{rxn}$, [m$^2$]) was computed based on regional chemical conversions

$$SA_{rxn} = l^2 \cdot \iiint_V \frac{q}{l_n^2} \cdot \frac{C_0 - C}{C_0} \cdot d\mathbf{r} \ . \qquad (7)$$

We also calculated the apparent surface area ($SA_{app}$, [m$^2$]) of a dissolving structure based on the instantaneous macroscopic mass balancing

$$SA_{app} = l^2 \cdot N \cdot \frac{l_n}{C_0} \cdot \Delta \varphi \qquad (8)$$

where $N$ is the number of voxels and $\Delta \varphi$ the change of the average porosity of the simulation domain over a time step. Note that Equation 8 has very often been used (improperly) to calculate reactive surface area.

The difference between the geometric and the reactive surface areas is demonstrated in Figure 1.

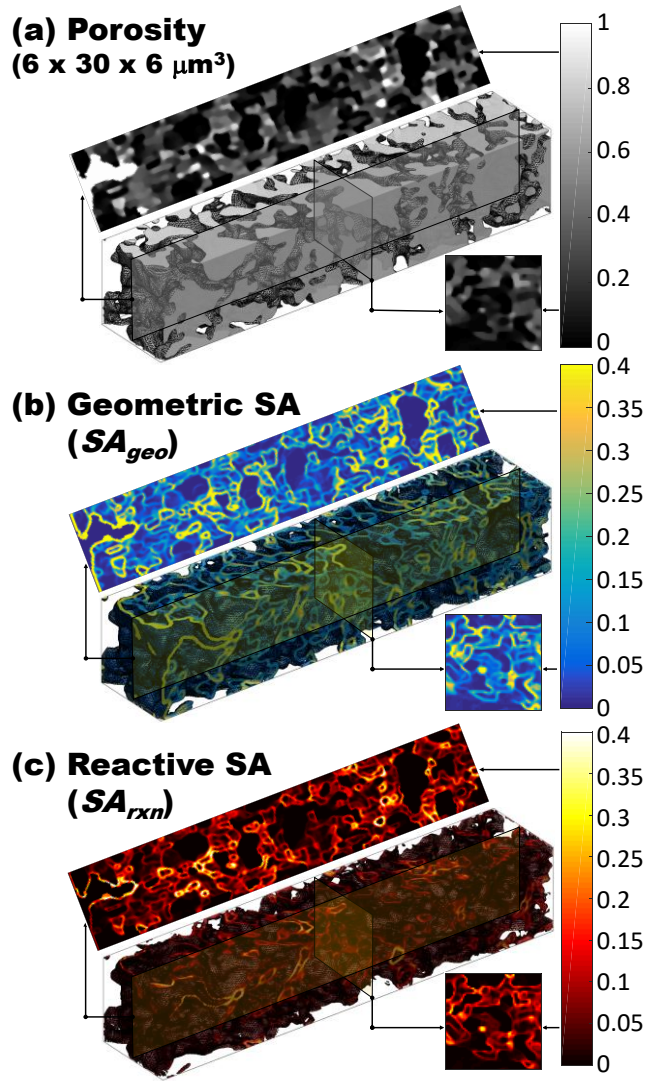

**Figure 1.** Geometric and reactive surface areas of a developing microstructure with a volume-averaged porosity of 0.237. (a) The developing microstructure based on grey scale porosity data. The isosurface was drawn at the median porosity of the dataset. Cross sections show grey levels at X = 30 (rectangular) and Y = 150 (square). (b) Distribution of geometric surface areas based on the spatial variations of material density (assuming mineralogical homogeneity). The magnitudes are normalised by the squared voxel size ($l^2 = 10^{-14}$ m$^2$). (c) Reactive surface area of the same microstructure. The isosurfaces in (b) and (c) were both drawn at the median of $SA_{geo}$ and scaled to 0.4 $l^2$ for comparison.

## RESULTS AND DISCUSSION

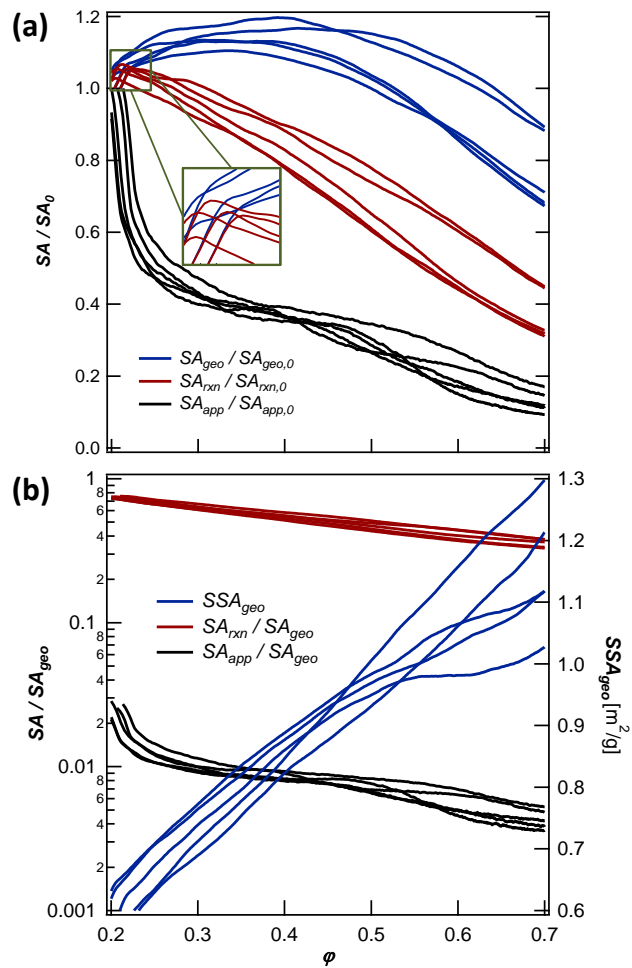

**Figure 2.** Evolution of surface area during microstructural dissolution. Curves of the same color show results from 5 different initial geometries. Geometric ($SA_{geo}$), reactive ($SA_{rxn}$) and apparent ($SA_{app}$) surface areas are plotted against the volume-averaged porosity ($\varphi$) as a measure of reaction progress. (a) Evolution of surface areas normalised to their initial values. The inset shows the initial concurrent increase of $SA_{rxn}$ with $SA_{geo}$ and the splitting of the curve pairs shortly after channel merging. (b) Discrepancies among differently defined surface areas. The specific surface area (SSA) shown was calculated based on $SA_{geo}$ and increased monotonically. Reactive and apparent surface areas were normalised to the geometric surface area of the same instant. $Q$ (normalized input flowrate) and $l_n$ (normalized voxel size) set to 1.

Figure 2 shows the evolution of the geometric ($SA_{geo}$), reactive ($SA_{rxn}$) and apparent ($SA_{app}$) surface areas of the dissolving samples. The geometric surface area reflects the spatial variations of material density[30] (Equation 6 and Figure 1b) and is independent of the volumetric flow rate $Q$ (normalized injection flowrate per voxel) or the dimensionless voxel size $l_n$. Curves of the same color show simulations using different initial geometries. The significance of the surface area is twofold: it not only serves as the interface for water-rock interactions, but also as a measure of transport heterogeneity that would be amplified by the reactive transport coupling (because $SA_{geo}$ is proportional to porosity differences among voxels). In all cases $SA_{geo}$ increased initially, reaching a maximum between $\varphi = 0.4$ and 0.5, and decreased until the depletion of solid material. This change in trend resulted from a counteraction between reactive infiltration instability (RII) and regional depletion of solid material. The effect of RII is explained in Figure 3: given the total flowrate ($Q$), the reactive fluid passing the two parallel channels A and B ($Q_A$ vs. $Q_B$) are distributed according to their relative permeabilities ($\kappa_A$ vs. $\kappa_B$). Now assume the fluid can modify the channels through a reversible chemical reaction whose rate is sensitive to the saturation state of the fluid. If $\kappa_A = \kappa_B$, then $Q_A$ and $Q_B$ will remain equal. If, however, there is a slight difference between $\kappa_A$ and $\kappa_B$ (say $\delta k = \kappa_B - \kappa_A > 0$), this difference will be amplified indefinitely with time because a fluid element passing through channel A would have resided in the channel shorter compared to one passing through channel B. This difference in residence time results in a lower saturation state at the exit (i.e., $C_A > C_B$). A less saturated fluid dissolves the solid faster, thus increasing $\kappa_A$ more rapidly than $\kappa_B$. This positive feedback will further decrease the residence time of fluid elements in channel A and, as a consequence, further increase the dissolution rate therein. Overall, this coupling between residence time and the rate of a reversible reaction through its chemical affinity amplifies any small transport heterogeneity ($\delta\kappa$) on its path unless the property is upper-bounded. For more elaborated mathematical treatments of RII, the readers are referred to the references.[1-3, 52]

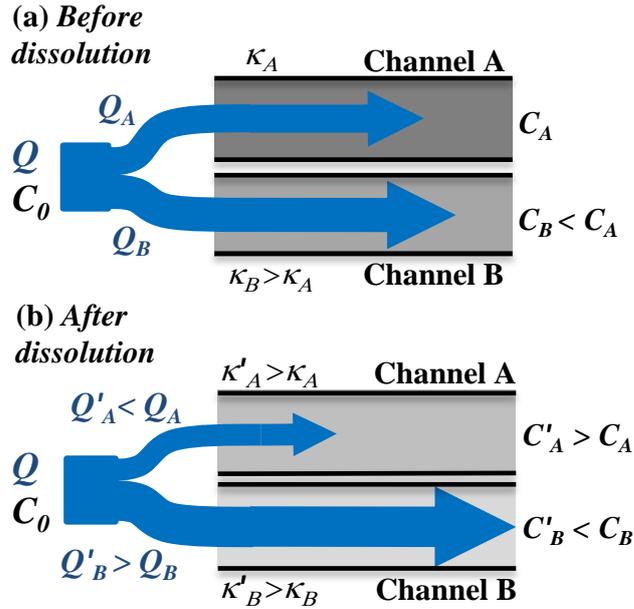

**Figure 3.** Reactive infiltration instability in dissolving media can amplify a very small difference in permeability indefinitely. (a) Initially the permeability of channel B was slightly greater than channel A. The outlet concentration of dissolved material $C_B$ was lower than $C_A$ because of a shorter fluid residence time. The media thus dissolved faster in channel B because the solution was less saturated. (b) The permeability of both channels increased. $\kappa_B$ increased faster because of the faster dissolution. Given the same global flowrate, $Q_A$ will decrease with an increasing $\kappa_A$.

Both transport and chemical heterogeneities are inherent to natural porous media. In this study we assumed that the media were chemically homogeneous and focused only on the transport heterogeneities stemming from a complex microstructure. The maximum number of heterogeneities in porous media characterized by tomography is proportional to the number of voxels (within each voxel the material is assumed homogeneous). The difference between the porosities of a neighbouring pair of voxels is proportional to the local geometric surface area. As a consequence, RII amplifies pre-existing geometric surface areas that have access to the reactive fluid, leading to an initial increase of $SA_{geo}$. Meanwhile, voxel porosity is upper-bounded at one. Inter-voxel porosity differences can be smeared out as dissolution removes the solid material in both neighbouring voxels completely. This depletion of solid ultimately leads to the disappearance of microstructure (as $\varphi \to 1$) and explains the decrease of $SA_{geo}$ over the course of the experiment. The evolution of $SA_{geo}$ is affected by the initial geometry and thus underpins the importance of a detailed characterisation of the microstructure.

The evolution of $SA_{geo}$ also depends on the macroscopic flowrate (Figure 4a). Higher flowrates lead to a higher maximum $SA_{geo}$ and a wider plateau. This dependency suggests that the structural morphing is controlled by access of reactive fluid to the geometric surface area. For example, Figure 5 shows the comparisons of geometric surface ($SA_{geo}$), reactive surface ($SA_{rxn}$), normalized regional fluid throughput ($q/Q$) and fluid reactivity $C = (C_{A,eq}-C_A)/(C_{A,eq}-C_{A,inj})$ between two simulations with identical initial geometry but different flowrates, where $C_A$, $C_{A,eq}$ and $C_{A,inj}$ represent concentration, equilibrium concentration and injecting concentration of the dissolution product. Given the same $C_{A,inj}$, a greater $Q$ brings more reactant into contact with pre-existing surfaces, allowing more heterogeneities to be amplified at a given instant. In case of fully segmented pores it can also be interpreted as by enlarging $Q$ the pressure gradient is higher thereby filling increasingly smaller pores with fluid. This wider spread of reactant resulted in a rapid initial increase of $SA_{geo}$. As the structure evolves, the same amount of removed solid (measured by $\Delta\varphi$) is distributed among more voxels in a system with a greater $Q$, and the solid would deplete slower regionally before the voxel porosities hit the cap and thus the plateau was maintained for a longer period. Note that accompanying the regional depletion of solid was the channelization of microstructure (wormholing)[31, 52] – a ubiquitous phenomenon leading to fluid "focusing". Figure 5 shows that the channelization of a smaller flow would take place within a lesser change of overall porosity ($\Delta\varphi$) because of the relatively confined distribution of aqueous reactant (Figure 5b: a smaller flowrate leads to more focused streams in $q/Q$, in concord with the narrower spread of $C$). The beginning of the plateau corresponded to the "merging" of multiple lesser channels near the fluid entrance into a major flow path, which then developed gradually towards the fluid outlet (discussed below). The plateau diminished when the developing major channel(s) reached the outlet ("breakthrough", see also Figure S5).

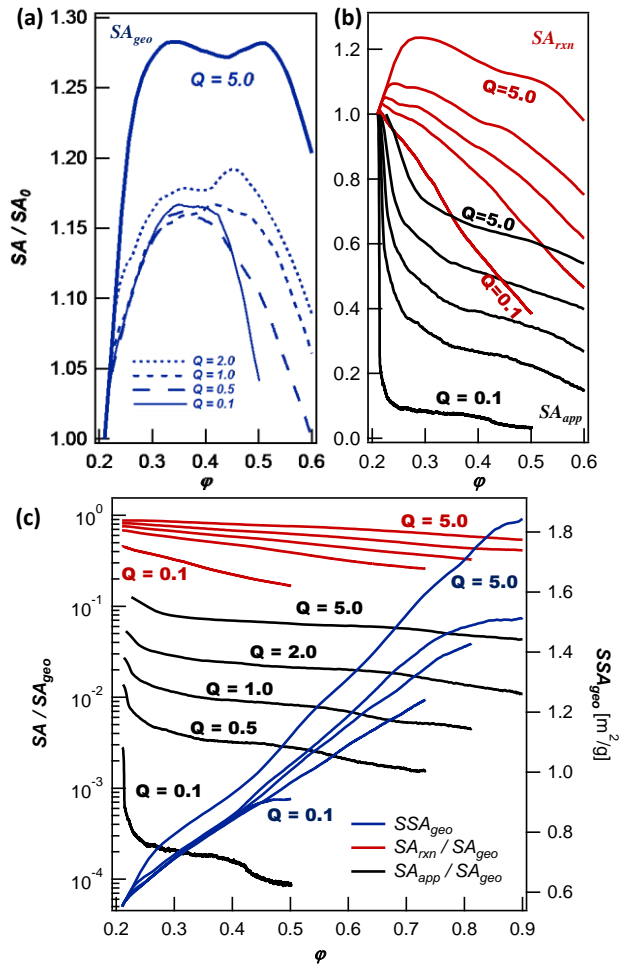

**Figure 4.** Flowrate dependency of surface area evolution. The same initial geometry was used in all simulations. $Q$ is the normalised injecting flowrate per voxel. $\varphi$ is the volume-averaged porosity of the simulation domain. (a) Evolution of geometric surface area. The plateaus during which the surface area was greater than the origin value are shown. Between $\varphi = 0.6$ and 1.0 the geometric surface areas decreased monotonically to zero. (b) Evolution of reactive (red) and apparent (black) surface areas. The results scaled with the macroscopic flowrate and thus only curves with the maximum and minimum flowrates are labelled. (c) Evolution of the differences between reactive (red)/apparent (black) and and the evolution of specific geometric surface area ($SSA_{geo}$) calculated from the density of calcite crystal (2.71 kg/L) (blue).

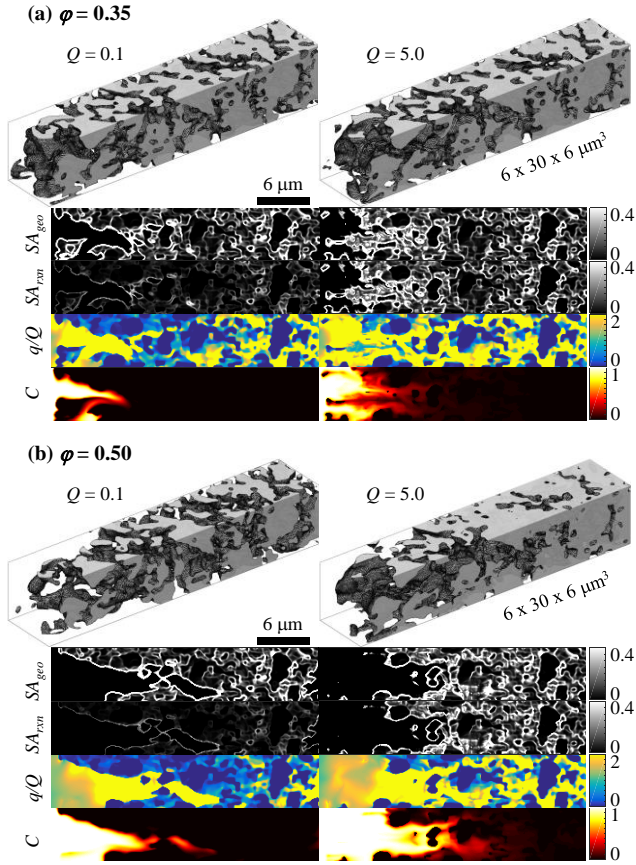

**Figure 5.** Effect of macroscopic flowrate on microstructural evolution in a 60x300x60 voxel domain. Cross sections depicture distributions of physical quantities at X = 30. The unit for both $SA_{geo}$ and $SA_{rxn}$ is the square of the voxel size ($l^2$), or $10^{-14}$ m$^2$. Distribution of regional fluid throughput is normalised to the macroscopic flowrate. The reactivity is normalised to the difference between the injection and the equilibrium concentrations (maximum reactivity available). At any instant, the greater flowrate leads to a wider spatial distribution of reactivity (*C*) and a less focused stream (*q/Q*). (a) and (b) show microstructures that evolved from the same initial geometry as the volume-averaged porosity reached 0.35 and 0.50.

At any given time, the reactive surface area ($SA_{rxn}$) is the portion of geometric surface area on which mineral dissolution takes place (Figure 1b). By definition (Equation 7), $SA_{rxn}$ is the spatially averaged chemical conversion $[(C-C_0)/C_0]$ weighted by the regional flowrate, *q*. The inclusion of $l_n$ and *q* suggests that $SA_{rxn}$ reflects the reactivities of both solid and fluid (in contrast to $SA_{geo}$ whose value is independent of the solution chemistry). Figure 3a shows that the initial evolution of $SA_{rxn}$ agreed well with the increase of $SA_{geo}$. This concurrence suggested that the new geometric surface areas created by instability of the system were fully utilized for dissolution. However, an abrupt change in the trend was observed in all simulations shortly before $SA_{geo}$ reaching the plateau, after which $SA_{rxn}$ decreased monotonically. This sudden decline was closely related to the initial morphological change near the fluid entrance. The increase in $SA_{rxn}$ requires a relatively homogeneous distribution of reactive fluid, so multiple small pores can develop in parallel (Figure 6a). This necessitates the absence of major preexisting fluid channels such as fractures in rocks and does only take place near the fluid entrance (where fluid is more reactive and multiple heterogeneities can be amplified in parallel). Once these pores are merged because of the complete dissolution of their "walls", fluid will be redirected into the major channel and no longer contribute to dissolution. The geometric surface area previously created by RII is cumulative in the calculation of $SA_{geo}$ but not in $SA_{rxn}$ because of this redistribution. This cumulative effect indicates that fluid focusing can be an important reason for the discrepancy between $SA_{geo}$ and $SA_{rxn}$, which will be amplified as dissolution progresses and channeling becomes more dramatic (Figure 5b). Meanwhile, the maximum of $SA_{rxn}$ signifies that the flow path(s) favored by the majority of fluid has been determined globally. Beyond this point the microstructural evolution is dominated by the expansion of these flow path(s). In Figure 5, for example, the shapes of the developing major flow paths are best visualized by the distribution of fluid reactivity $C$.

The evolution of $SA_{rxn}$ shows strong flowrate dependence (Figure 4b) for two reasons. First, a greater amount of reactant injection initiates the development of more pores in parallel, thus a faster increase of $SA_{geo}$, which can be fully utilized for dissolution before major channel forms. Second, channeling a lesser stream (a smaller volumetric flowrate) required a smaller conduit. This difference in the responses to the material removal can be clearly seen in Figure 5, where given an increase of 0.15 in the macroscopic porosity ($\varphi$) the smaller stream ($Q$ = 0.1) was "narrower" distributed ($q/Q$) and reached out farther towards the outlet. The $SA_{rxn}$ then decreased rapidly because the majority of reactant bypassed the solid material upstream through the small conduit. The discrepancy between the geometric and the reactive surface areas thus depended on macroscopic flowrate, with a smaller $Q$ leading to a greater discrepancy. It is noteworthy that this discrepancy did not show high sensitivity to the initial geometry of a chemically homogeneous system in this study (Figure 2b).

Surface area has also been defined based on macroscopic chemical conversion (Equation 9). Practically, this conversion can be obtained from a comparison between the composition of solutions at the inlet and outlet,[9] or from the rate of solid material disappearence.[53, 54] Such defined surface areas have frequently been considered as reactive surface area. In this study we refer to this definition as the apparent surface area ($SA_{app}$) and show that in porous media exhibiting infiltration instability, $SA_{app}$ and $SA_{rxn}$ may show significantly different dynamics. Figure 2a shows the evolution of $SA_{app}$ for different initial geometries. In contrast to $SA_{geo}$ and $SA_{rxn}$, $SA_{app}$ does not include any geometric information in its definition (i.e., it does not require the spatial variation of $\varphi$ as an input). It decreased monotonically, and its evolution can be roughly divided into three stages. The first stage was a rapid drop of $SA_{app}$. This drop is usually referred to as the "fast initial dissolution" of rocks.[44] Fast initial dissolution has been observed in both percolation systems (e.g., core flooding or column experiments)[55] and stirred systems (batch or continuous flow experiments).[40] This "abnormality" of mineral dissolution has been attributed to two reasons.[56] First, a higher reactive surface area preexists and depletes rapidly as dissolution begins. Second, the geometric surface area can be more reactive initially because of the mechanical truncation of chemical bonds (e.g., through grinding). However, our results were obtained from chemically homogeneous materials (all surfaces were equally reactive) whose geometric surface areas were initially increasing (instead of depleting), and thus ruled out both explanations.

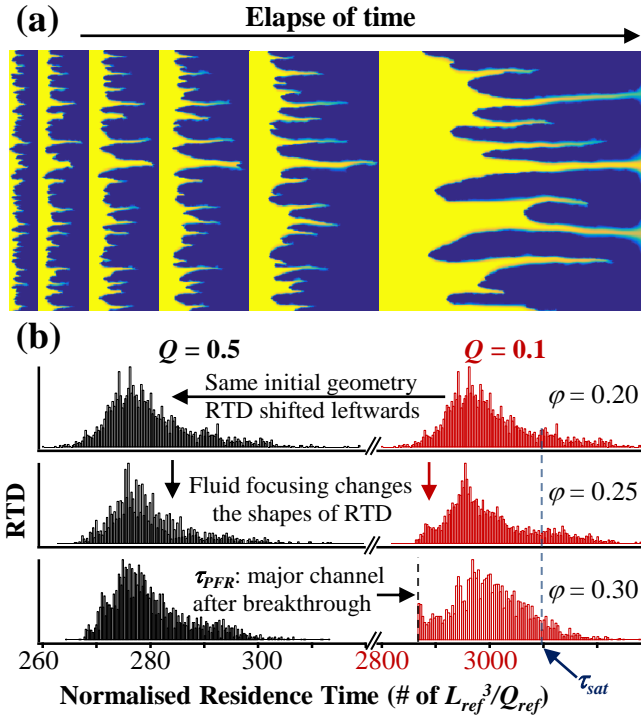

**Figure 6.** Fluid focusing and its effect on Residence Time Distribution (RTD). (a) 2D demonstration of the parallel development and merging of microchannels. Reactive fluid flowing from left to right. Color represents the normalized concentration of reactant, with yellow and blue being 1 and 0. At any given instant, the decrease rate of chemical conversion is proportional to the number of developing channels. This number reduces over time because porosity is capped at 1 and small channels merge. As a result, the chemical conversion (proportional to $SA_{app}$) decreases most rapidly at the beginning of a percolation. (b) Effects of fluid focusing and flowrate on RTD. Given the same microstructure, varying flowrate shifts an RTD horizontally without changing its shape. Fluid focusing, in contrast, deforms an RTD by moving its bulk leftwards. This deformation decreases the mean residence time. After breakthrough, an RTD is left-bounded by the residence time of fluid in the major channel ($\tau_{PFR}$). This channel serves as a hollow plug flow reactor where dissolution takes place only on the wall. Note that given the same amount of dissolved solid ($\Delta \varphi = 0.10$ in figure b), breakthrough took place first with the smaller flowrate ($Q = 0.1$), consistent with the narrower flow channels observed morphologically (Figure 5). The residence time for the reactive fluid to reach equilibrium ($\tau_{sat}$) is critical because only fluid elements to its left can effectively use geometric surface for reaction. Here $\tau_{sat}$ is shown for demonstration purpose. In general, $\tau_{sat}$ is not a constant in porous media because of the uneven distribution of surface area along each flow path.

The fast initial dissolution of the microstructure, as well as the rapid decrease of $SA_{app}$, are results of fluid focusing. Fluid focusing is the consequence of channel formation during which smaller streams combine, reducing the number of flow paths while increasing the superficial flowrate. This focusing process can effectively reduce the residence time of reactants. Figure 6b shows the effect of fluid focusing on the residence time distribution (RTD) of fluid. The RTD was calculated by first computing the streamlines originated from each of the voxels at the fluid inlet and then integrating the reciprocal of regional velocity along the streamlines. In general, chemical conversion increases with residence time.[50] Fluid focusing can "gather" reactants into bigger flow channels, shortening the residence time and "drags" its mean leftwards. As a result, the rapid decrease of $SA_{app}$ reflects the progress of fluid focusing, and its rate is proportional to the number of developing microchannels (Figure 6a). In our simulations, the number of simultaneously developing pores was maximized at the beginning of a percolation, and quickly dropped before entering the plateau stage of $SA_{geo}$. This dynamics suggested that the decrease of $SA_{app}$ should have occurred with the initial merging of upstream pores during which the reactive surface area ($SA_{rxn}$) actually increased. Furthermore, the rate of decrease was lowered after the merge to reflect the growth of the predominant channel(s) towards the fluid outlet. A second inflexion may appear to signify the channel breakthrough, after which the evolution of $SA_{app}$ was governed by the expanding of the major flow channel. The dissolution patterns during the three stages of $SA_{app}$ evolution are summarized in Figure S5.

The effects of macroscopic flowrate ($Q$) on $SA_{app}$ were manifold. Given the same microstructure, a greater $Q$ shifts the RTD leftwards without changing its shape (Figure 6b). This shift reduces chemical conversion per fluid volume and decreases $SA_{app}$. However, as discussed above, larger $Q$ also brings more reactant into the system, effectively increasing the wide spread of reaction front (Figure 5b). Because $SA_{app}$ is calculated assuming implicitly that the whole volume of a porous medium is dissolving (e.g., in Equation 9 the conversion is normalized to the full domain with 1.08M voxels instead of the voxels containing the active flow paths only), this increase in reactant consumption increases the macroscopic chemical conversion. The dependence of dissolution rate on reactant concentration further complicates the effect of $Q$. Greater $Q$ reduces conversion and increases the dissolution rate on average because the fluid is farther away from equilibrium. In contrast, a small and more focused stream gets equilibrated quickly without reaching out for more geometric surface area. In Figure 6b, $\tau_{sat}$ denotes the residence time a fluid element needed to achieve "virtually" complete conversion (a complete conversion indicates equilibrium. With the first order kinetics scheme used in this study, the reactive fluid can approach equilibrium indefinitely but can never reach equilibrium). The relative position of an RTD to $\tau_{sat}$ affects the macroscopic conversion (and thus $SA_{app}$) significantly because only the fluid with a residence time to the left of $\tau_{sat}$ can use all geometric surfaces in contact before leaving the medium. Overall, the effect of fluid reactivity (the amount of reactant input) was dominant and $SA_{app}$ scaled roughly with $Q$. It is worth mentioning that very often the difference between $SA_{app}$ and $SA_{geo}$ is referred to as the "discrepancy between reactive and geometric surface area".[21] Figure 4c shows that this difference ($SA_{app}$ and $SA_{geo}$) is even more pronounced than the real difference between the two surface areas ($SA_{rxn}$ and $SA_{geo}$) because of the opposite initial trends ($SA_{geo}$ increased with decreasing $SA_{app}$).

In summary, the simulations suggested that differences among geometric, reactive and apparent surface areas were amplified in a pressure driven process with dissolving minerals. These differences scaled inversely with the macroscopic flowrate. Morphological analysis showed that the channelization of the pore structure and the subsequent focusing of reactive fluid were the main reasons for the observed discrepancies, and may trigger a fast initial dissolution even if the material is chemically homogeneous. We hope this study can be part of clarifying the long lasting ambiguity in the conceptualization of reactive surface area and thus lay the foundation for reconciling inconsistencies among kinetics measurements of water-rock interactions.

ASSOCIATED CONTENT

**Supporting Information**. Additional discussion on earliness of mixing (EOM), dimensional analysis (DA) and 5 supporting figures.

AUTHOR INFORMATION

**Corresponding Author**

*E-mail: yiyang@nano.ku.dk

**Notes**

The authors declare no competing financial interest.

ACKNOWLEDGMENT

Funding for this project was provided by the Innovation Fund Denmark through the CINEMA project, the Innovation Fund Denmark and Maersk Oil and Gas A/S through the P$^3$ project as well as the European Union's Horizon 2020 research and innovation programme under the Marie Sklodowska-Curie grant agreement No 653241. Support for synchrotron beamtime was received from the Danish Council for Independent Research via DANSCATT. We thank Heikki Suhonen at the ID22 beamline at ESRF (The European Synchrotron) for technical support.

ABBREVIATIONS

DA, dimensional analysis; EOM, earliness of mixing; EOR, enhanced oil recovery; GCS, geologic carbon storage; MMM, maximum mixedness model; RII, reactive infiltration instability; SFM, segregated flow model.

TOC Figure

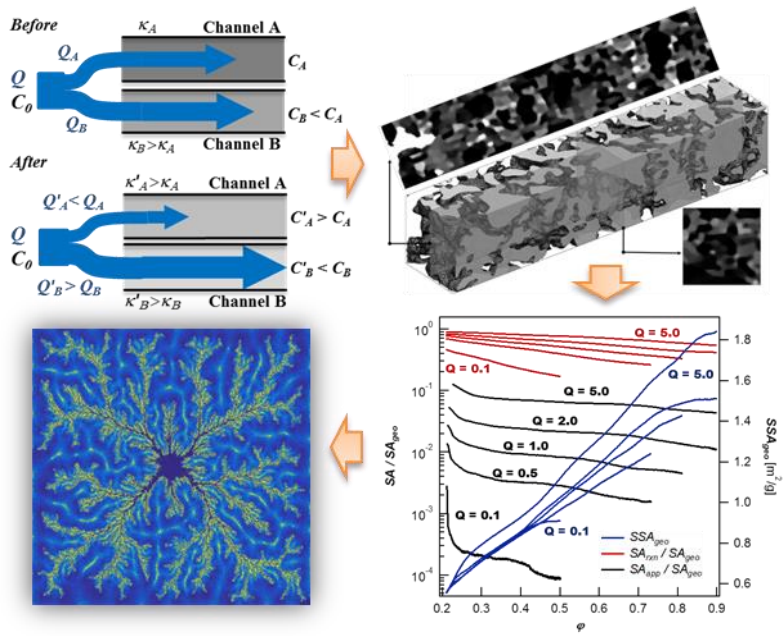

REFERENCES

1. Ortoleva, P.; Chadam, J.; Merino, E.; Sen, A., Geochemical self-organization II: the reactive-infiltration instability. *Am. J. Sci* **1987,** *287*, 1008-1040.
2. Szymczak, P.; Ladd, A. J. C., Reactive-infiltration instabilities in rocks. Fracture dissolution. *Journal of Fluid Mechanics* **2012,** *702*, 239-264.
3. Chadam, J.; Hoff, D.; Merino, E.; Ortoleva, P.; Sen, A., Reactive Infiltration Instabilities. *IMA Journal of Applied Mathematics* **1986,** *36*, (3), 207-221.
4. Deng, H.; Fitts, J. P.; Crandall, D.; McIntyre, D.; Peters, C. A., Alterations of Fractures in Carbonate Rocks by CO2-Acidified Brines. *Environmental Science & Technology* **2015,** *49*, (16), 10226-10234.
5. Deng, H.; Ellis, B. R.; Peters, C. A.; Fitts, J. P.; Crandall, D.; Bromhal, G. S., Modifications of Carbonate Fracture Hydrodynamic Properties by CO2-Acidified Brine Flow. *Energy & Fuels* **2013,** *27*, (8), 4221-4231.
6. Olsson, J.; Bovet, N.; Makovicky, E.; Bechgaard, K.; Balogh, Z.; Stipp, S. L. S., Olivine reactivity with CO2 and H2O on a microscale: Implications for carbon sequestration. *Geochimica et Cosmochimica Acta* **2012,** *77*, 86-97.
7. Noiriel, C.; Steefel, C. I.; Yang, L.; Ajo-Franklin, J., Upscaling calcium carbonate precipitation rates from pore to continuum scale. *Chemical Geology* **2012,** *318–319*, 60-74.
8. Gouze, P.; Luquot, L., X-ray microtomography characterization of porosity, permeability and reactive surface changes during dissolution. *Journal of Contaminant Hydrology* **2011,** *120–121*, 45-55.
9. Noiriel, C.; Luquot, L.; Madé, B.; Raimbault, L.; Gouze, P.; van der Lee, J., Changes in reactive surface area during limestone dissolution: An experimental and modelling study. *Chemical Geology* **2009,** *265*, (1–2), 160-170.
10. Luquot, L.; Gouze, P., Experimental determination of porosity and permeability changes induced by injection of CO2 into carbonate rocks. *Chemical Geology* **2009,** *265*, (1–2), 148-159.
11. Ellis, B. R.; Fitts, J. P.; Bromhal, G. S.; McIntyre, D. L.; Tappero, R.; Peters, C. A., Dissolution-driven permeability reduction of a fractured carbonate caprock. *Environmental engineering science* **2013,** *30*, (4), 187-193.
12. Fitts, J. P.; Peters, C. A., Caprock Fracture Dissolution and CO2 Leakage. *Reviews in Mineralogy and Geochemistry* **2013,** *77*, (1), 459-479.
13. Hubbert, M. K.; Willis, D. G., Mechanics of hydraulic fracturing. **1972**.
14. Schechter, R. S., Oil well stimulation. **1992**.
15. Chadam, J.; Ortoleva, P.; Sen, A., A weakly nonlinear stability analysis of the reactive infiltration interface. *SIAM Journal on Applied Mathematics* **1988,** *48*, (6), 1362-1378.
16. Røyne, A.; Jamtveit, B., Pore-Scale Controls on Reaction-Driven Fracturing. *Reviews in Mineralogy and Geochemistry* **2015,** *80*, (1), 25-44.
17. Wildenschild, D.; Sheppard, A. P., X-ray imaging and analysis techniques for quantifying pore-scale structure and processes in subsurface porous medium systems. *Advances in Water Resources* **2013,** *51*, 217-246.
18. Zachara, J.; Brantley, S.; Chorover, J.; Ewing, R.; Kerisit, S.; Liu, C.; Perfect, E.; Rother, G.; Stack, A. G., Internal Domains of Natural Porous Media Revealed: Critical Locations for Transport, Storage, and Chemical Reaction. *Environmental Science & Technology* **2016,** *50*, (6), 2811-2829.
19. Noiriel, C., Resolving Time-dependent Evolution of Pore-Scale Structure, Permeability and Reactivity using X-ray Microtomography. *Reviews in Mineralogy and Geochemistry* **2015,** *80*, (1), 247-285.

20. Yoon, H.; Kang, Q.; Valocchi, A. J., Lattice Boltzmann-Based Approaches for Pore-Scale Reactive Transport. *Reviews in Mineralogy and Geochemistry* **2015,** *80*, (1), 393-431.
21. Steefel, C. I.; Beckingham, L. E.; Landrot, G., Micro-Continuum Approaches for Modeling Pore-Scale Geochemical Processes. *Reviews in Mineralogy and Geochemistry* **2015,** *80*, (1), 217-246.
22. Molins, S., Reactive Interfaces in Direct Numerical Simulation of Pore-Scale Processes. *Reviews in Mineralogy and Geochemistry* **2015,** *80*, (1), 461-481.
23. Mehmani, Y.; Balhoff, M. T., Mesoscale and Hybrid Models of Fluid Flow and Solute Transport. *Reviews in Mineralogy and Geochemistry* **2015,** *80*, (1), 433-459.
24. Molins, S.; Trebotich, D.; Steefel, C. I.; Shen, C., An investigation of the effect of pore scale flow on average geochemical reaction rates using direct numerical simulation. *Water Resources Research* **2012,** *48*, (3).
25. Schlüter, S.; Sheppard, A.; Brown, K.; Wildenschild, D., Image processing of multiphase images obtained via X-ray microtomography: A review. *Water Resources Research* **2014,** *50*, (4), 3615-3639.
26. Andrä, H.; Combaret, N.; Dvorkin, J.; Glatt, E.; Han, J.; Kabel, M.; Keehm, Y.; Krzikalla, F.; Lee, M.; Madonna, C.; Marsh, M.; Mukerji, T.; Saenger, E. H.; Sain, R.; Saxena, N.; Ricker, S.; Wiegmann, A.; Zhan, X., Digital rock physics benchmarks—Part I: Imaging and segmentation. *Computers & Geosciences* **2013,** *50*, 25-32.
27. Müter, D.; Pedersen, S.; Sørensen, H. O.; Feidenhans'l, R.; Stipp, S. L. S., Improved segmentation of X-ray tomography data from porous rocks using a dual filtering approach. *Computers & Geosciences* **2012,** *49*, 131-139.
28. R., G.; S., B.; D., M.; A., M.; P., H. R.; S, S. S. L.; O., S. H., Effect of Tomography Resolution on the Calculated Microscopic Properties of Porous Media: Comparison of Carbonate and Sandstone Rocks. *Applied Physics Letters (under review)* **2016**.
29. Liu, C.; Liu, Y.; Kerisit, S.; Zachara, J., Pore-Scale Process Coupling and Effective Surface Reaction Rates in Heterogeneous Subsurface Materials. *Reviews in Mineralogy and Geochemistry* **2015,** *80*, (1), 191-216.
30. Yeong, C. L. Y.; Torquato, S., Reconstructing random media. *Physical Review E* **1998,** *57*, (1), 495-506.
31. Steefel, C. I.; Lasaga, A. C., A coupled model for transport of multiple chemical-species and kinetic precipitation dissolution reactions with application to reactive flow in single-phase hydrothermal systems. *American Journal of science* **1994**, (294), 529-592.
32. White, A. F.; Brantley, S. L., The effect of time on the weathering of silicate minerals: why do weathering rates differ in the laboratory and field? *Chemical Geology* **2003,** *202*, (3), 479-506.
33. Bear, J., *Dynamics of fluids in porous media*. Courier Corporation: 2013.
34. Landrot, G.; Ajo-Franklin, J. B.; Yang, L.; Cabrini, S.; Steefel, C. I., Measurement of accessible reactive surface area in a sandstone, with application to CO 2 mineralization. *Chemical Geology* **2012,** *318*, 113-125.
35. Malmström, M. E.; Destouni, G.; Banwart, S. A.; Strömberg, B. H., Resolving the scale-dependence of mineral weathering rates. *Environmental science & technology* **2000,** *34*, (7), 1375-1378.
36. Fredd, C. N.; Fogler, H. S., Influence of transport and reaction on wormhole formation in porous media. *AIChE Journal* **1998,** *44*, (9), 1933-1949.
37. Zhu, C.; Lu, P.; Zheng, Z.; Ganor, J., Coupled alkali feldspar dissolution and secondary mineral precipitation in batch systems: 4. Numerical modeling of kinetic reaction paths. *Geochimica et Cosmochimica Acta* **2010,** *74*, (14), 3963-3983.

38. Maher, K.; Steefel, C. I.; DePaolo, D. J.; Viani, B. E., The mineral dissolution rate conundrum: Insights from reactive transport modeling of U isotopes and pore fluid chemistry in marine sediments. *Geochimica et Cosmochimica Acta* **2006,** *70*, (2), 337-363.
39. Yang, Y.; Min, Y.; Lococo, J.; Jun, Y.-S., Effects of Al/Si ordering on feldspar dissolution: Part I. Crystallographic control on the stoichiometry of dissolution reaction. *Geochimica et Cosmochimica Acta* **2014,** *126*, 574-594.
40. Yang, Y.; Min, Y.; Jun, Y.-S., Effects of Al/Si ordering on feldspar dissolution: Part II. The pH dependence of plagioclases' dissolution rates. *Geochimica et Cosmochimica Acta* **2014,** *126*, 595-613.
41. Montazeri, M.; Moreau, J.; Uldall, A.; Nielsen, L. In *Full waveform seismic modelling of Chalk Group rocks from the Danish North Sea-implications for velocity analysis*, EGU General Assembly Conference Abstracts, 2015; 2015; p 9513.
42. Muter, D.; Sorensen, H.; Jha, D.; Harti, R.; Dalby, K.; Suhonen, H.; Feidenhans'l, R.; Engstrom, F.; Stipp, S., Resolution dependence of petrophysical parameters derived from X-ray tomography of chalk. *Applied Physics Letters* **2014,** *105*, (4).
43. Zhu, W.; Liu, J.; Elsworth, D.; Polak, A.; Grader, A.; Sheng, J.; Liu, J., Tracer transport in a fractured chalk: X-ray CT characterization and digital-image-based (DIB) simulation. *Transport in Porous Media* **2007,** *70*, (1), 25-42.
44. Brantley, S. L.; Kubicki, J. D.; White, A. F., *Kinetics of water-rock interaction*. Springer: 2008; Vol. 168.
45. Cloetens, P.; Ludwig, W.; Baruchel, J.; Van Dyck, D.; Van Landuyt, J.; Guigay, J.; Schlenker, M., Holotomography: Quantitative phase tomography with micrometer resolution using hard synchrotron radiation x rays. *Applied physics letters* **1999,** *75*, (19), 2912-2914.
46. Bruns, S.; Stipp, S. L. S.; Sørensen, H. O., Looking for the Signal: A Guide to Iterative Noise and Artefact Removal in X-ray Tomography Reconstructions of Reservoir Rocks. *Water Resources Research (under review)* **2016**.
47. Bruns, S.; Sørensen, H. O.; Stipp, S. L. S., Rock Properties of Compacted North Sea Chalks characterized by Greyscale Analysis. *Water Resources Research (under review)* **2016**.
48. Jha, D.; Sørensen, H. O.; Dobberschütz, S.; Feidenhans; apos; l, R.; Stipp, S. L. S., Adaptive center determination for effective suppression of ring artifacts in tomography images. *Applied Physics Letters* **2014,** *105*, (14), 143107.
49. Fogler, H. S., Elements of chemical reaction engineering. **1999**.
50. Levenspiel, O., *Chemical reaction engineering*. Wiley & Sons, Inc.: 1999.
51. Lasaga, A. C., *Kinetic theory in the earth sciences*. Princeton University Press: 2014.
52. Steefel, C. I.; Lasaga, A. C. In *Evolution of dissolution patterns: Permeability change due to coupled flow and reaction*, ACS symposium series, 1990; Oxford University Press: 1990; pp 212-225.
53. Rufe, E.; Hochella, M. F., Quantitative Assessment of Reactive Surface Area of Phlogopite During Acid Dissolution. *Science* **1999,** *285*, (5429), 874-876.
54. Smith, M. M.; Sholokhova, Y.; Hao, Y.; Carroll, S. A., Evaporite caprock integrity: An experimental study of reactive mineralogy and pore-scale heterogeneity during brine-CO2 exposure. *Environmental science & technology* **2012,** *47*, (1), 262-268.
55. Olsson, J.; Stipp, S. L. S.; Dalby, K. N.; Gislason, S. R., Rapid release of metal salts and nutrients from the 2011 Grímsvötn, Iceland volcanic ash. *Geochimica et Cosmochimica Acta* **2013,** *123*, 134-149.
56. Parsons, I., *Feldspars and their Reactions*. Springer Science & Business Media: 2012; Vol. 421.

# Supporting Information: Reactive infiltration instability amplifies the difference between geometric and reactive surface areas in natural porous materials

Y.Yang*, S. Bruns, S. L. S. Stipp, H. O. Sørensen

 Nano-Science Center, Department of Chemistry, University of Copenhagen, Universitetsparken 5, DK-2100 Copenhagen, Denmark

**Earliness of Mixing**

Computing the chemical conversion of a reactant requires the knowledge of contacting pattern in a reactor. This knowledge consists of two types of information: residence time distribution (RTD) and earliness of mixing (EOM). RTD reflects the macroscopic mixing state and its effect on chemical conversion depends on the monotonicity of a reaction rate law, i.e. the first derivative of reaction rate with respect to reactant concentration. When this derivative equals zero, RTD does not affect conversion. EOM reflects the molecular mixing of materials and its effect on conversion depends on the concavity of the rate law, i.e. the second derivative of reaction rate with respect to reactant concentration. Together, RTD and EOM quantify the mixing state of a reactor, which represents the "net effect" of momentum and mass transfer. For example, a well stirred vessel with turbulence and a stagnant cup of hot water with strong diffusion both show good internal mixing, despite dramatic differences in transport mechanisms. The reaction scheme chosen in this study, $A_{solid} \rightleftharpoons A_{solute}$ with $r_A = k_A \left( C_{A,eq} - C_A \right)$, is concavity free and thus avoided the need to quantify EOM beyond the resolution limit (sub-voxel mixing). Figure S4 shows the comprehensive design of the reactor network model where the quantification of both RTD and EOM are necessary on the resolvable and unresolvable structures. In this scheme, Segregated Flow Model (SFM) and Maximum Mixedness Model (MMM) as well as a few other microscopic isothermal mass transfer models can be used to quantify mixing within each voxel.

**Dimensional Analysis**

Dimensional analysis (DA) was carried out to minimize the number of parameters in the mathematical scheme. DA is a practice of systematically lumping parameters together by redefining the base units of physical quantities. This is allowed because the selection of units does not affect the physical laws governing system evolution. In our simulation, three dimensionless groups may affect the system dynamics:

$$Rz = \left(\frac{P_{ref} L_{ref}}{\mu Q_{ref}}\right) \cdot \left(\frac{\partial \kappa}{\partial \varphi}\right),$$

$$Hn = \left(\frac{L_{ref}^2}{Q_{ref}}\right) \cdot \left|\frac{\partial r_A}{\partial C_A}\right| \text{ and}$$

$$Ds = \left(\frac{Q_{ref}}{L_{ref}^3}\right) \cdot \left(\frac{M}{\rho}\right) \cdot (C_{A,eq} - C_{A,inj}) \cdot dt$$

where $\mu$ is the viscosity of the reactive fluid [Pa·s], $\varphi$ is the porosity of a voxel, $l_n$ is the normalized voxel size, $\frac{\partial \kappa}{\partial \varphi}$ is the first order Taylor coefficient of voxel level permeability [m$^2$], $r_A$ is the mineral dissolution rate [mol·m$^{-2}$·s$^{-1}$], $\frac{\partial r_A}{\partial C_A}$ is the first order rate constant [m·s$^{-1}$], $C_{A,eq}$ is the equilibrium concentration of dissolved solid [mol·m$^{-3}$], $C_A$ is the concentration of dissolved $A$ [mol·m$^{-3}$], $C_0$ and $C$ are normalized inlet and outlet reactant concentrations, $M$ and $\rho$ are the molar weight [g·mol$^{-1}$] and the density [g·L$^{-1}$] of the mineral, $C_{A,inj}$ is the concentration of dissolved $A$ in the macroscopic injection fluid and $P_{ref}$, $L_{ref}$ and $Q_{ref}$ are reference pressure (Pa), length [m] and volumetric flowrate [m$^3$/s], respectively. These three dimensionless groups represent the strength of coupling between different physical and chemical processes. $Rz$ reflects the sensitivity of flow field to microstructure, $Hn$ reflects the sensitivity of kinetics to aqueous composition and $Ds$ reflects the sensitivity of microstructure change to chemical reactions. Should any of these dimensionless groups be zero, the positive feedback that leads to infiltration instability vanishes and microstructural evolution halts. Further note that the inclusion of three reference values of physical quantities ($Q_{ref}$, $P_{ref}$ and $L_{ref}$) guaranteed the existence of a base unit combination that would make all dimensionless groups equal to one (or any specified set of real numbers). This result suggests that regardless what values do the involved physical and chemical parameters take – as long as they are not zero – they only rescale the size of the simulation domain by changing the value of the normalized voxel size $l_n$. In this study, we chose the base units to be

$$L_{ref} = \frac{(C_{A,eq} - C_{A,inj})dt}{Ds} \cdot \left(\frac{M}{\rho}\right) \cdot \left(\frac{|\partial r_A/\partial C_A|}{Hn}\right),$$

$$P_{ref} = \left(\frac{Rz \cdot \mu}{\partial \kappa/\partial \varphi}\right) \cdot \frac{(C_{A,eq} - C_{A,inj})dt}{Ds} \cdot \left(\frac{M}{\rho}\right) \cdot \left(\frac{\partial r_A/\partial C_A}{Hn}\right)^2 \text{ and}$$

$$Q_{ref} = \left[\frac{(C_{A,eq} - C_{A,inj})dt}{Ds} \cdot \left(\frac{M}{\rho}\right)\right]^2 \cdot \left(\frac{|\partial r_A/\partial C_A|}{Hn}\right)^3.$$

Hence, the governing equations reduced to

$$q = l_n \varphi^2 \cdot \Delta p,$$

$$-C_0 + \exp\left[(1-\eta) \cdot l_n^2 \cdot \left(\frac{\Delta \varphi}{q}\right)\right] \cdot C = 0,$$

$$-\sum_i q_i C_{0,i} + \left[1 + \eta \cdot l_n^2 \cdot \left(\frac{\sum_{i=1}^6 |\nabla \varphi|_i}{q}\right)\right] \cdot qC = 0 \text{ and}$$

$$d\varphi = l_n^{-3} \cdot \sum_{i=1}^7 q_i (C_{0,i} - C_i).$$

And the inputs for a numerical simulation become

1) Macroscopic flowrate and fluid composition
2) Microstructure ($\varphi$ as a function of space, from tomography)
3) Dimensionless voxel size $l_n$

4) Sub-voxel mixing factor $\eta$

in which (1) reflects the interactions between the system and the environment, (2) and (3) are intrinsic properties of a system and (4) is a fitting parameter.

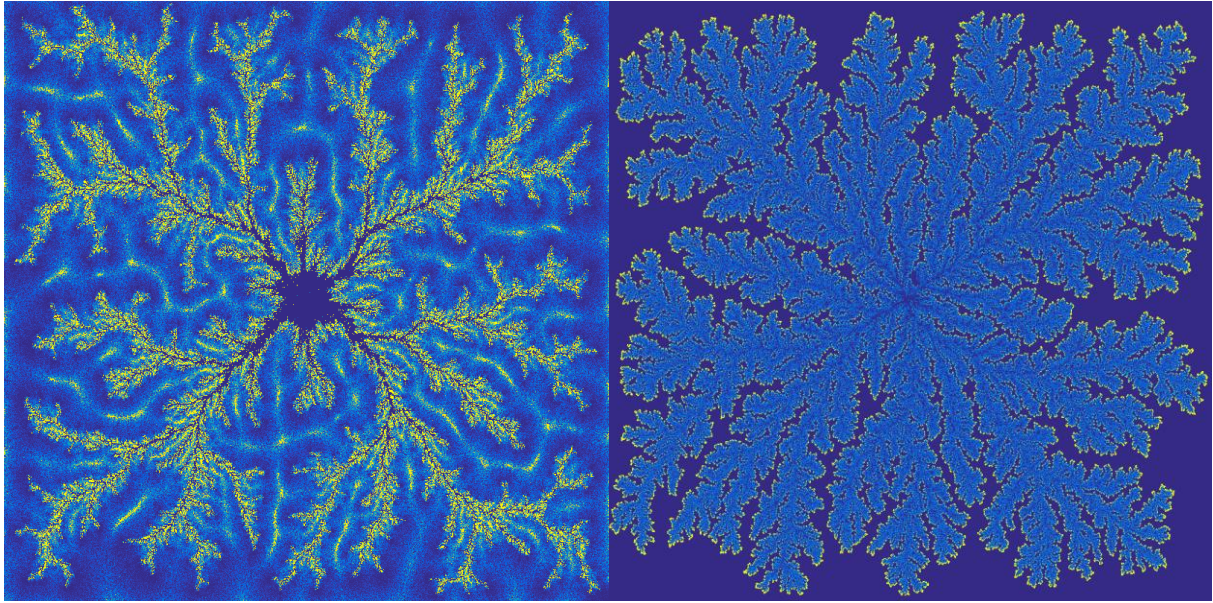

**Figure S1.** Morphological examples for a simulated 2D flow system development induced by infiltration instability. Left: regional transport property unbounded. Right: regional transport property upper bounded. Figures show the morphological development of a 1001 by 1001 pixels domain after 1000 time steps. The reactive fluid was injected at the central pixel (X=Y=501) and allowed to migrate freely outwards. A uniformly distributed random variation in the transport property, amounting to 20% of the mean, was applied to the initial domain as regional heterogeneities.

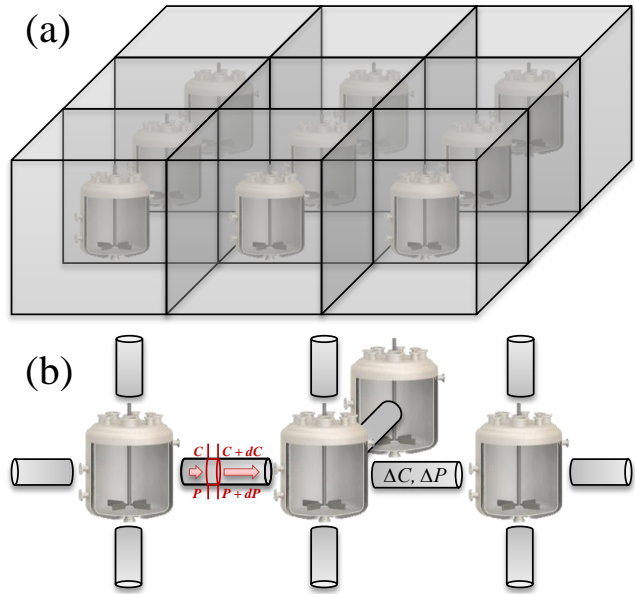

**Figure S2.** Schematic of the reactor network model. Every reactor amounts to voxel in the tomography image. (a) Voxels are modelled as continuously stirred tank reactors (CSTRs). (b) Mass and momentum transfer between neighboring voxels through plug flow reactors (PFRs). The inlet and outlet of each reactor is determined by the transient macroscopic pressure field.

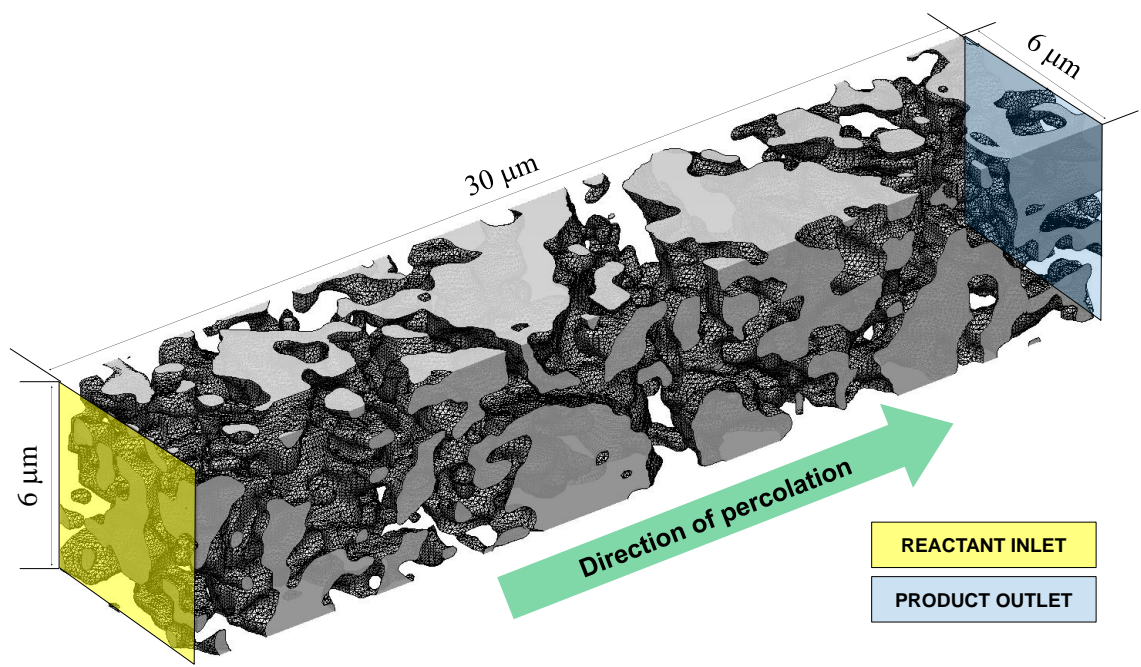

**Figure S3.** Schematic representation of the simulation setup. The domain consisted of 1.08 million voxels representing randomly selected 6 x 30 x 6 µm³ subvolumes of the chalk sample. Reactive fluid was injected evenly from the 3600 voxels at Y = 0 (reactant inlet) and removed from the 3600 voxels at Y = 300 (product outlet). The porosity, permeability and surface area of each reactor was parameterised according to the signal intensity of the host voxel. Isosurface drawn at the mean porosity of the grey scale dataset ($\varphi = 0.21$).

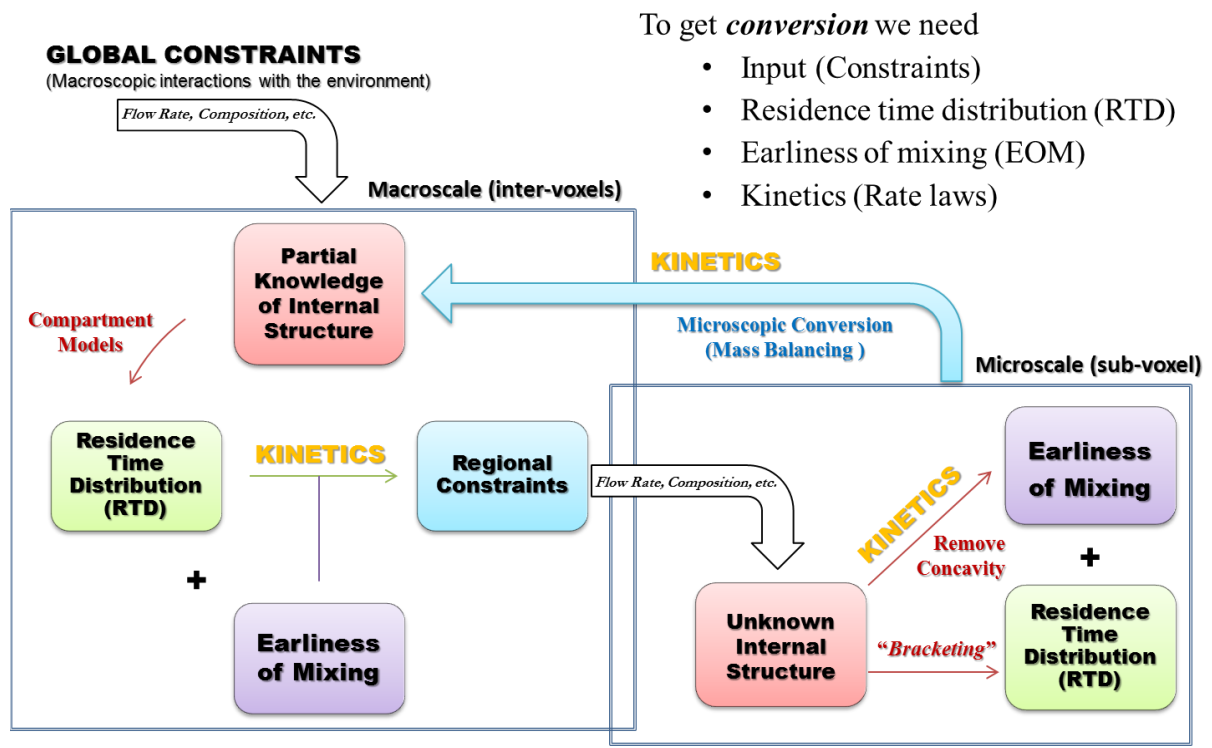

**Figure S4.** Overview of the reactor network model. The nested design shows the different strategies of quantifying contact patterns at different scales. The voxel size represents the boundary between resolvable and unresolvable microstructures. On the macroscale, Residence Time Distribution (RTD) can be estimated based on flow field while Earliness Of Mixing (EOM) can be bracketed with Maximum Mixedness Model (MMM) and Segregated Flow Model (SFM). On the sub-voxel level, microscopic transport models parameterised through complementary characterisation techniques can be used to estimate both RTD and EOM.

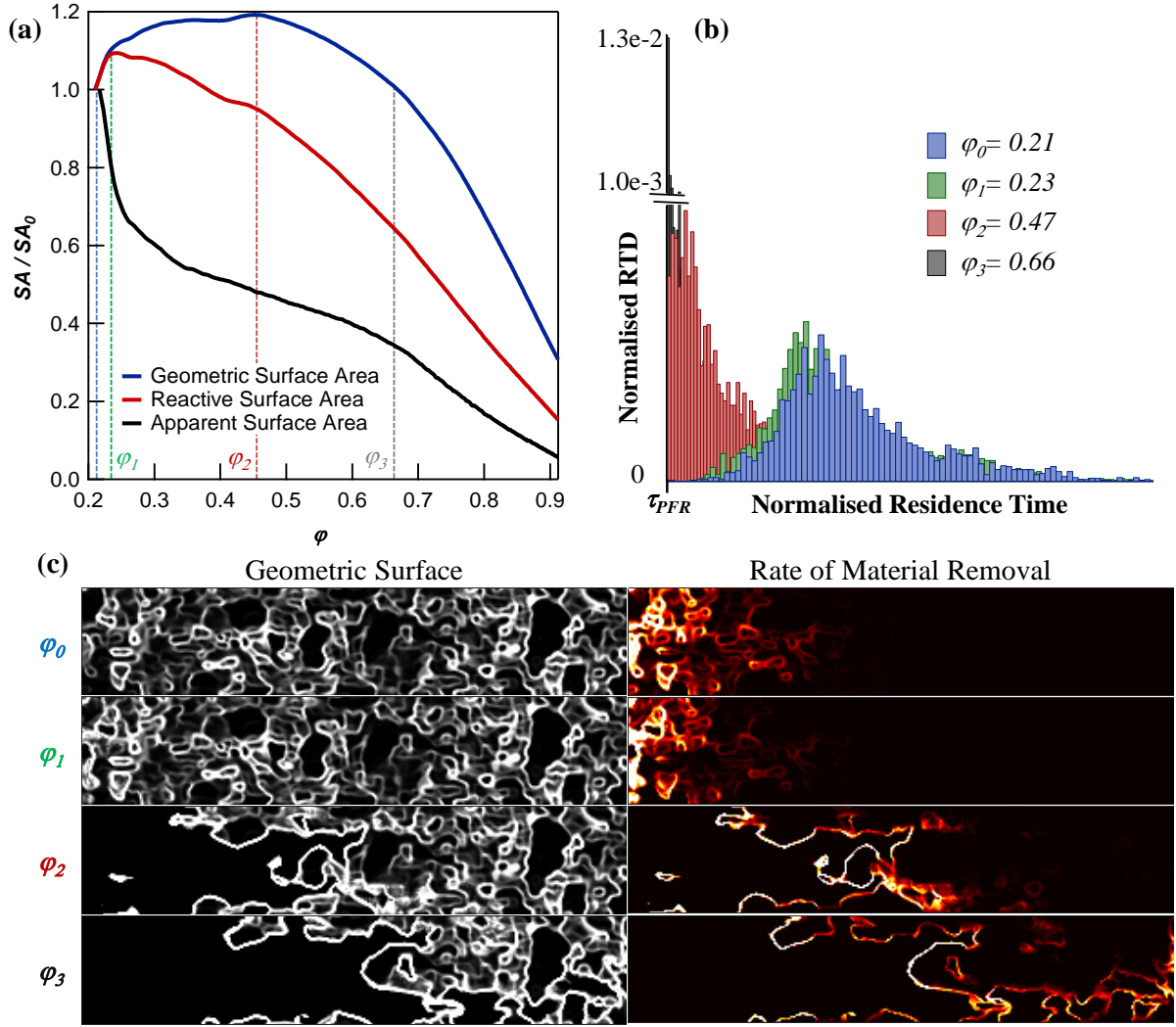

**Figure S5.** Dissolution patterns during different stages of microstructural evolution ($Q = 2.0$). (a) Evolution of surface areas. $\varphi_1$ marks the end of the initial merging of micropores and the beginning of major channel development $\varphi_2$ is the porosity when the increase of surface area by infiltration instability was cancelled out by the decrease of solid availability. $\varphi_3$ shows the breakthrough porosity after which the structure development was dominated by the expansion of the major channel. (b) Effect of fluid focusing on residence time distribution (RTD). Fluid focusing leads to the decrease of both residence time and solid-liquid contact and is the main reason for the decrease of $SA_{app}$. (c) Geometric surface area (left) and the distribution of material removal rate (right). The dissolution was relatively homogeneous initially because no preferential flow path existed and reactant was distributed evenly. Later the dissolution became "surface controlled" because fluid focusing made upstream geometric surface unavailable to reactants.